# Direct transformation of crystalline MoO$_3$ into few-layers MoS$_2$

Felix Carrascoso *[1], Gabriel Sanchez-Santolino †,[1], Chun-wei Hsu [1,2], Norbert M. Nemes [3], Almudena Torres-Pardo [4], Patricia Gant [1], Federico J. Mompeán [1], Kourosh Kalantar-zadeh [5], José A. Alonso [1], Mar García-Hernández [1], Riccardo Frisenda [1], Andres Castellanos-Gomez *[1]

[1] Materials Science Factory. Instituto de Ciencia de Materiales de Madrid (ICMM-CSIC), Madrid, E-28049, Spain.
[2] Kavli Institute of Nanoscience. Delft University of Technology. Delft 2600 GA, The Netherlands.
[3] Departamento de Física de Materiales, Universidad Complutense de Madrid, E-28040 Madrid, Spain.
[4] Departamento de Química Inorgánica, Facultad de Químicas, Universidad Complutense, 28040- Madrid, Spain.
[5] School of Chemical Engineering, University of New South Wales, Kensington, NSW 2052, Australia.
* Correspondence: felix.c@csic.es , andres.castellanos@csic.es

**Abstract:** We fabricate large-area atomically thin MoS$_2$ layers through the direct transformation of crystalline molybdenum trioxide (MoO$_3$) by sulfurization at relatively low temperatures. The obtained MoS$_2$ sheets are polycrystalline (~10-20 nm single-crystal domain size) with areas of up to 300×300 μm$^2$ with 2-4 layers in thickness and show a marked p-type behaviour. The synthesized films are characterized by a combination of complementary techniques: Raman spectroscopy, X-ray diffraction, transmission electron microscopy and electronic transport measurements.

## 1. Introduction

Two-dimensional (2D) transition metal dichalcogenides (TMDCs) have recently gained interest among the scientific community to solve the weakness of the lack of a bandgap in graphene, which limits its applications in field-effect transistors and digital integrated circuits [1]. The TMDC molybdenum disulphide (MoS$_2$) was the first 2D material with an intrinsic bandgap that was isolated [2] and it consists of S-Mo-S layers that are held by weak van der Waal forces in a trigonal prismatic structure [3–6]. In its bulk form, this material displays an indirect bandgap of about 1.2 eV; nevertheless, it becomes a direct bandgap semiconductor (1.8 eV) when it is thinned down to a monolayer [7]. In addition, when a single-layer MoS$_2$ is used as the channel in a field-effect transistor, it exhibits high in-plane mobility and large current ON/OFF ratio [8]. These are the reasons why molybdenum disulphide has attracted interest for electronic and optoelectronics applications [8–10]. Furthermore, it is an attractive candidate for energy conversion [11,12] and storage [13,14], hydrogen evolution reactions [15–17] or oxygen reduction reactions [18].

First methods that were reported for the synthesis of 2D MoS$_2$ consisted of mechanical and chemical exfoliation from bulk crystals [2,19,20] and, in fact, a lot of studies still use these methods since they provide high-quality single layers. However, these techniques present some problems like randomly deposited flakes, relatively small coverage area of material and a poor control over thickness. A solution for these issues is critical to achieve real-life electronic devices based on MoS$_2$ and, therefore, synthesis of large-area MoS$_2$ films is a very active research area. The most explored methods to synthesise large-area MoS$_2$ thin films are the chemical vapour deposition (CVD) [21,22] and the sulfuration of sputtered molybdenum thin films [23–25].

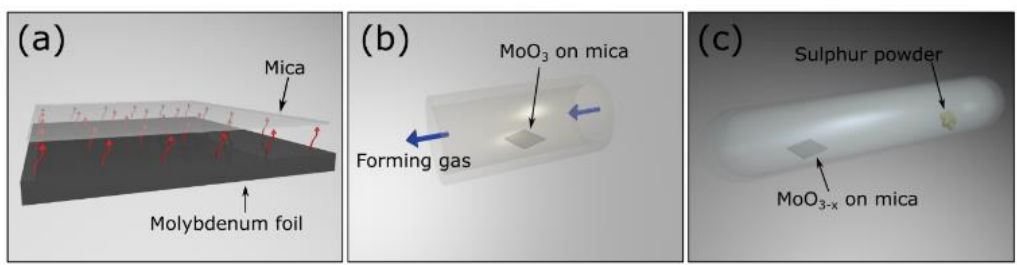

**Figure 1.** Cartoon of the process followed for the MoO$_3$ conversion into MoS$_2$. **(a)** MoO$_3$ sublimes from a hot molybdenum foil (540 °C) and it crystallizes onto a mica substrate. **(b)** MoO$_{3-x}$ is formed after placing the MoO$_3$ in a tube furnace at 300 °C in a forming gas atmosphere for 24 hours. **(c)** Sulfuration process is performed in a closed glass ampoule at 500 °C - 600 °C.

Here, we explore an alternative route to obtain atomically thin MoS$_2$ layers: the direct transformation of crystalline molybdenum trioxide (MoO$_3$) layers into MoS$_2$ nanosheets by sulfurization at moderate temperatures. Up to now the sulfurization of crystalline MoO$_3$ have only demonstrated to produce MoS$_2$ fullerenes and nanotubes but here we demonstrate that it can be also employed to fabricate large area MoS$_2$ layers [26,27]. We characterized the resulting layers by Raman spectroscopy, X-ray diffraction and transmission electron microscopy, finding that the resulting layers show all the characteristics of polycrystalline MoS$_2$. We transferred the as-synthesized films to pre-patterned electrodes to fabricate electronic devices and we found that they are strongly p-doped, which can be an interesting feature to complement the marked n-doping of mechanically exfoliated or CVD grown MoS$_2$. Our synthesis method does not require a tube furnace with flow gas control as the sulfurization is carried out in a sealed ampoule, simplifying considerably its implementation, and reducing the cost.

**2. Materials and Methods**

The crystalline MoO$_3$ source is obtained by heating up a molybdenum foil (99.99% purity) to 540 °C in air using a laboratory hot plate. At this temperature, the MoO$_3$ starts to sublime. A mica substrate is placed above the hot molybdenum foil. The MoO$_3$ gas sublimed from the hot molybdenum foil crystalizes on the slightly cooler mica substrate placed on top, as we show in Figure 1(a). As reported by *Molina-Mendoza et al.* [26], this method produces continuous crystalline thin films through a van der Waals epitaxy process thanks to the van der Waals interaction with the mica surface. Note that in the van der Waals epitaxy process there is no need for lattice matching between the substrate and the grown MoO$_3$ overlayer.

Prior to the sulfuration of the MoO$_3$ crystals, they are reduced by heating them at 300 °C for 24 hours in a tube furnace in forming gas atmosphere, Figure 1(b). This process yields MoO$_{3-x}$ crystals. We found that this step is crucial to avoid the evaporation of MoO$_3$ during the sulfuration process as MoO$_3$ is a highly volatile material. On the contrary, MoO$_2$ is a more stable oxide [28], in fact, by partially reducing the molybdenum trioxide we observe an improved stability of the material upon temperature increase. Then the MoO$_{3-x}$ layers are converted to MoS$_2$ by a sulfuration process in a closed glass ampoule. The sample containing the MoO$_{3-x}$ layers is sealed with sulphur powder at 10$^{-5}$ mbar pressure. The ampoule is placed in a furnace at 500 °C for 5 hours and then the temperature is increased at 600 °C for another 5 hours. Once the sulfuration process is concluded, the temperature is slowly lowered to room temperature, Fig 1(c). The number of MoS$_2$ layers that we obtain depends on the starting $MoO_3$ thickness. Therefore, with this method we are able to obtain $MoS_2$ continuous layers covering most of the mica substrate with regions of up to 300×300 $\mu m^2$ with <5 layers in thickness, an example is shown in Figure 2(a). As discussed below, also single layer $MoS_2$ could be observed (see the discussion related to the scanning transmission electron microscopy results). It is important to note that, when we tried to sulfurize the as-grown $MoO_3$ layers, without the reduction step, we obtained thick $MoS_2$ crystallites randomly deposited on both the ampoule surface and on the substrate. Scanning transmission electron microscopy (STEM) data was acquired in an aberration-corrected JEOL JEM-ARM200cF electron microscope operated at 80 kV.

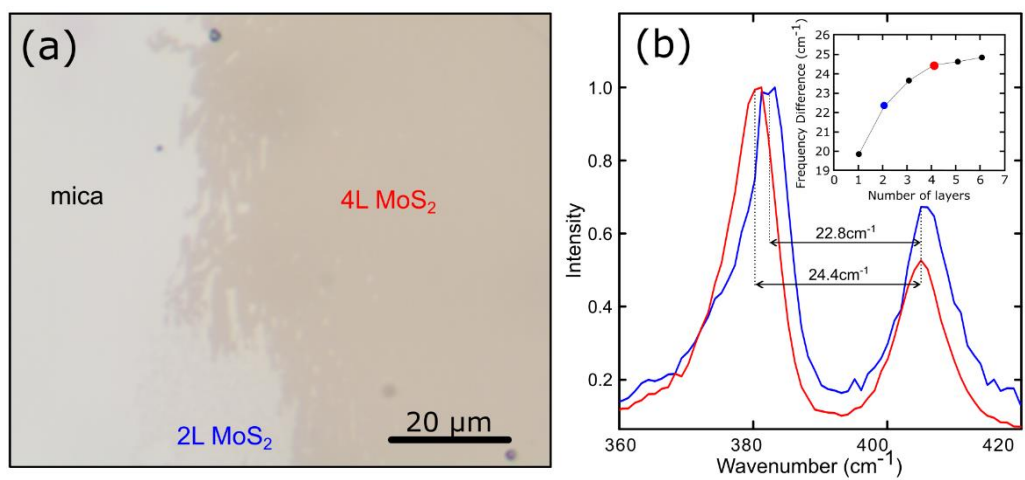

**Figure 2. (a)** Optical image of a large area $MoS_2$ on a mica substrate. **(b)** Raman Spectra of $MoS_2$ in different regions of the same sample. The inset displays the relation between the frequency difference of the two peaks and the number of layers of $MoS_2$.

## 3. Results and discussion

*3.1. Raman characterization*

In Figure 2(a) we show an optical image of a thin and large-area $MoS_2$ film on mica. We employ Raman spectroscopy to characterize the $MoS_2$ film as this technique has been demonstrated to be a very powerful tool to characterize 2D materials [29,30]. Figure 2(b) presents the Raman spectra acquired on two locations (indicated in the figure) of the $MoS_2$ film shown in Figure 2(a). The characteristic $E^1_{2g}$ and $A_{1g}$ phonon modes of $MoS_2$ (around 380 $cm^{-1}$ and 415 $cm^{-1}$) are clearly visible in the spectra [23,31]. One can determine the number of layers from the frequency difference between these two Raman modes. In the inset in Figure 2(b) we show the relation between this frequency difference and the number of layers of $MoS_2$, obtained from the literature [32,33], and we compare these values with those obtained in two spots in our sample to determine the number of layers finding that the $MoS_2$ specimen is composed of a bilayer and a four-layer region. We address the reader to the Supporting Information for a Raman map of another thin $MoS_2$ region.

*3.2. XRD characterization*

The crystal structure of the films has been characterized with X-ray diffraction (XRD). XRD was performed at room temperature on the initial sample ($MoO_3$ grown on mica 18 mm x 2 mm substrate) and on the final sample ($MoS_2$ obtained after the sulfuration process). Figure 3 illustrates the X-ray diffractograms that were taken for the initial sample and for the final sample in green and blue, respectively. In red, we also show the X-ray diffractogram for a bare mica substrate in order to be able to differentiate the peaks that belong to the substrate from the peaks that correspond to the growth film.

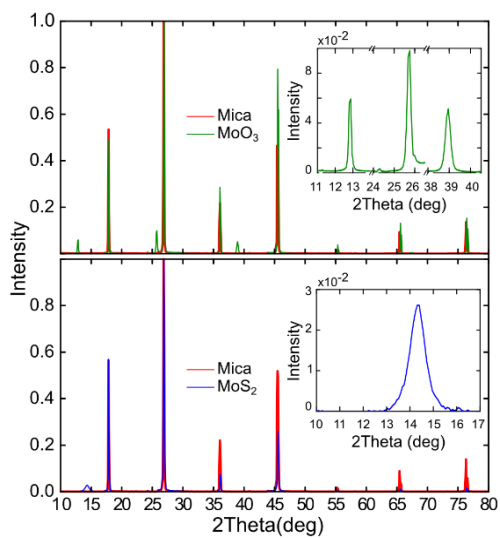

**Figure 3.** Comparison of XRD spectra of the sample shown in Fig. 2, at different steps: MoO$_3$ (initial) and MoS$_2$ (after sulfuration) in green and blue, respectively. XRD spectra of a mica substrate in red to distinguish it from the peaks of the layer analysed (insets).

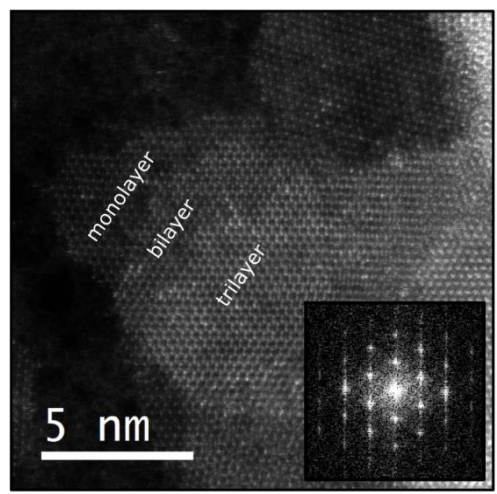

**Figure 4.** High magnification HAADF images of a MoS$_2$ thin film transferred over a holey Si$_3$N$_4$ membrane support. Inset shows FFT where a clearly hexagonal symmetry is exhibited.

Notice that the green spectrum exhibits peaks that correspond to (0 2 0), (0 4 0) and (0 6 0) reflections, which belong to the diffraction peaks of MoO$_3$. The appearance of the (0 k 0) peaks, parallel to the plane (0 1 0), is a product of a preferred orientation of the MoO$_3$ crystal with respect to the mica (0 0 1) surface due to the van der Waals epitaxy type of growth [34]. The blue spectrum obtained for the same sample after the sulfuration process shows a peak that corresponds to the (0 0 2) reflection of MoS$_2$ [6]. Thus, we further confirm that we are able to obtain MoS$_2$ from MoO$_3$ deposited onto a mica substrate. In some works it is proposed that the average thickness of a thin sample can be obtained from the analysis of the XRD peaks using the Scherrer equation ($D=k\lambda/\beta cos\theta$, where $k$ is the shape factor, $\lambda$ is the X-ray wavelength, $\beta$ is the full width at half maximum of the peak and $2\theta$ is the scattering angle) [35,36]. By analysing the (002) peak of the MoS$_2$ XRD pattern we estimated a c-stacking height for the analyzed sample of 10 nm, which corresponds to 15 layers of MoS$_2$. Note that this value corresponds to the average thickness of the whole sample; however thinner regions (as those shown in Figure 2) can be found on it. It is also worth mentioning that the single-crystal domain size observed in our samples is also of the order of ~10 nm (see STEM discussion below) and thus it is not completely clear if the Scherrer equation provides accurate values of the average thickness of the sample or simply the single-crystal domain size.

*3.2. STEM characterization*

The crystal structure of the films can be further characterized in real space by STEM. Figure 4 displays a high-angle annular dark field (HAADF) image of a MoS$_2$ layer transferred over a holey Si$_3$N$_4$ membrane support by an all-dry deterministic transfer process [37]. In order to transfer the MoS$_2$ films on mica we stick a polydimethylsiloxane (PDMS) sheet on its surface and we immerse it in distilled water. Due to the hydrophilic character of mica, the water wedges between de MoS$_2$ and the mica surface separating the MoS$_2$ layer, which keeps attached to the PDMS substrate, from the mica surface. The MoS$_2$ is easily transferred to the membrane by gently pressing the PDMS containing the MoS$_2$ film against the acceptor substrate and peeling it off slowly.

The STEM characterization indicates that the MoS$_2$ film is polycrystalline with a single-crystal domain size of 10-20 nm. Thinner regions can be found at the edges of the sulfurized film, where one can find monolayer, bilayer and trilayer areas (Figure 4 shows the edge of a MoS$_2$ film where mono-, bi- and tri-layer areas can be resolved). The fast Fourier transform (FFT) obtained from the monolayer region shows clearly the hexagonal symmetry of MoS$_2$.

*3.2. Electrical characterization*

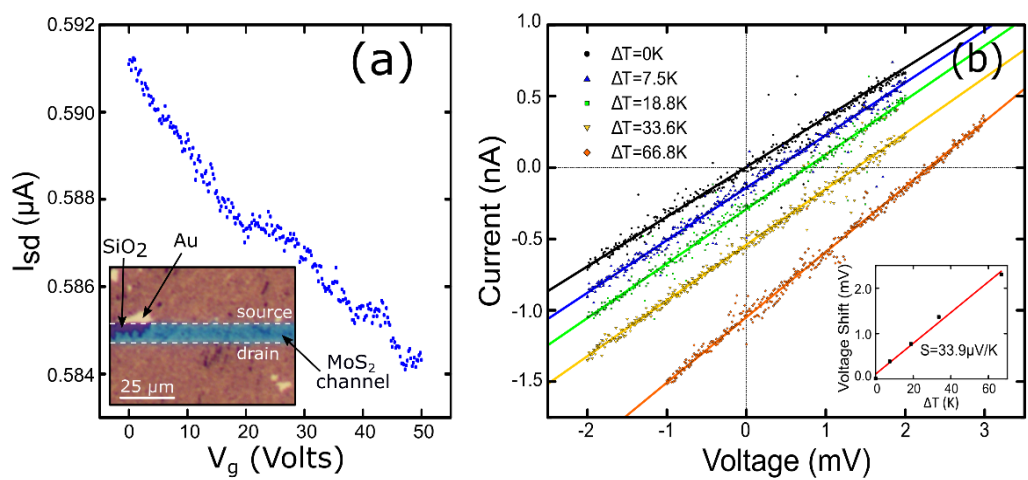

**Figure 5.** (a) Source-drain current *vs*. gate voltage measured in dark conditions and at $V_{sd}$ = 1 V. The inset shows an optical image of the device measured (channel length = 10 μm, channel width = 1 mm). (b) Seebeck effect measurement on a MoS$_2$ layer on a mica substrate by applying a temperature difference between electrodes. The inset shows the linear relationship between the thermovoltage shift and the difference in temperature. The Seebeck coefficient can be readily extracted from the slope.

The electrical properties of the fabricated MoS$_2$ films have been characterized fabricating a field-effect device, by transferring a MoS$_2$ film onto a SiO$_2$/Si with pre-patterned drain-source electrodes separated 10 μm. Figure 5(a) shows the measured source-drain current *vs*. gate voltage ($I_{sd}$-$V_g$) characteristic for a fixed source-drain voltage of $V_{sd}$ = 1 V. Surprisingly, we obtain a decrease in the source-drain current upon gate voltage increases without reaching the OFF state, which corresponds to a strong p-doped field effect behaviour. To confirm this fact, we performed a thermopower measurement. Figure 5(b) displays the *IV* characteristics acquired applying a temperature difference between the two electrodes. It can be seen how a positive voltage offset at zero current appears (thermoelectric voltage) when the temperature different increases. The inset shows the thermoelectric voltage versus the temperature difference. The Seebeck coefficient can be extracted from the slope of a linear fit to the data: $S$ = + 33.9 μV/K. This positive value confirms the p-doped nature of the MoS$_2$ film obtained by the direct sulfurization of crystalline MoO$_3$. The low magnitude of the Seebeck coefficient also indicates a high doping level We have carried out preliminary Hall effect measurements backing up the p-type electrical behavior of the MoS$_2$ films observed in the Seebeck and electric-field measurements. Unfortunately, the large resistance of our samples precludes us from quantifying the charge carrier concentration as the electronics of our Hall effect measuring system is optimized for low impedance samples. The highly linear shape of the *IV*s, together with the high doping inferred from the shallow transconductance and low Seebeck coefficient, points to an Ohmic contact in the Au-MoS$_2$ junction. We have also estimated the resistivity of the device ~100 Ω·cm, which is significantly higher to that of single-crystal $MoS_2$ (~1-5 Ω·cm),[38,39] as expected from the small single-crystal domain size of our synthetic $MoS_2$ layers.

In order to get a deeper insight into the microscopic origin of this p-doping in our $MoS_2$ layers we have done an electron energy loss spectra (EELS) analysis of the STEM data (see Supporting Information). Apart from the presence of Mo and S, we found C (which could come from e-beam induced deposition of amorphous carbon during the STEM measurement), O and B. The presence of O could be due to an incomplete $MoO_3$ to $MoS_2$ transformation and the presence of B impurities could come from unintentional cross-contamination from the surface of the glass ampoules used during the growth. The presence of these foreign species could be a plausible source of the unexpected p-type doping.

Figure 6(a) represents the measured $I_{sd}$-$V_{sd}$ characteristics in dark condition and under light excitation with different wavelengths. The gate voltage was set to $V_g = 0$ V during the measurement. Fiber coupled LED light sources were employed to illuminate the device. The inset of this figure zooms on the high voltage region of the traces to distinguish the differences induced upon illumination. The photocurrent as a function of the wavelength can be calculated from these data, as we show in Figure 6(b). This spectrum reveals that the maximum photocurrent value is located between 530 nm and 595 nm, whereas it decreases at longer wavelengths. We were not able of measuring a sizeable photocurrent beyond 740 nm as expected for multilayer $MoS_2$.

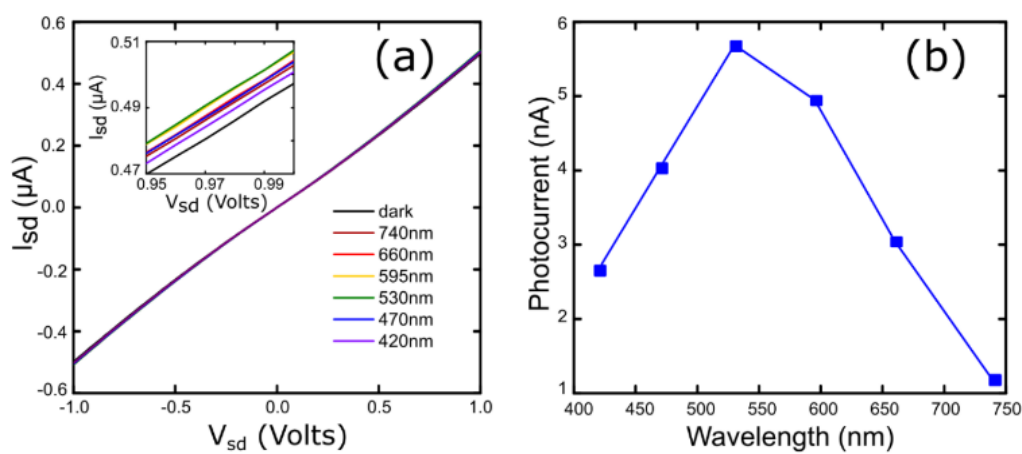

**Figure 6. (a)** $I_{sd}$-$V_{sd}$ curves for different illumination wavelengths and $V_g = 0$ V. Inset shows a smaller range to facilitate the visualization of the generated photocurrent. **(b)** Photocurrent spectrum obtained from $I_{sd}$-$V_{sd}$ curves.

## 4. Conclusions

In summary, we presented an alternative method to obtain atomically thin $MoS_2$ layers through the direct transformation of crystalline molybdenum trioxide ($MoO_3$) layers into $MoS_2$ nanosheets by sulfurization. The process can be carried out at moderate temperatures and using simple instrumentation. We obtained large area polycrystalline $MoS_2$ sheets 2 to 4 layers thick and we characterized them by Raman spectroscopy, X-ray diffraction and transmission electron microscopy. Regarding their electronic properties, they are strongly p-doped.

**Funding:** ACG acknowledge funding from the European Research Council (ERC) under the European Union's Horizon 2020 research and innovation programme (grant agreement n° 755655, ERC-StG 2017 project 2D-TOPSENSE) and from the EU Graphene Flagship funding (Grant Graphene Core 2, 785219). JAA thanks the Spanish Ministry of Economy, Industry and Competitiveness (MINECO) for funding the project MAT2017-84496-R. RF acknowledges support from MINECO through a "Juan de la Cierva: formación" fellowship (2017 FJCI-2017-32919). GSS acknowledges financial support from MINECO (Juan de la Cierva 2015 program, FJCI-2015-25427).

**Acknowledgments:** We thank Carmen Munuera (ICMM-CSIC) and the National Center for Electron Microscopy (CNME; UCM, Madrid) for facilities.

**Author Contibutions:** Conceptualization, Andres Castellanos-Gomez; Formal analysis, Felix Carrascoso, Gabriel Sanchez-Santolino, Chun-wei Hsu , Almudena Torres-Pardo , Federico Mompean and Riccardo Frisenda; Funding acquisition, Andres Castellanos-Gomez; Investigation, Felix Carrascoso, Gabriel Sanchez-Santolino, Chun-wei Hsu , Norbert M. Nemes , Almudena Torres-Pardo , Patricia Gant, Federico Mompean, Riccardo Frisenda and Andres Castellanos-Gomez; Project administration, Andres Castellanos-Gomez; Resources, Jose A. Alonso, Mar Garcian-Hernandez and Andres Castellanos-Gomez; Supervision, Norbert M. Nemes , Jose A. Alonso, Mar Garcian-Hernandez, Riccardo Frisenda and Andres Castellanos-Gomez; Writing – original draft, Felix Carrascoso and Andres Castellanos-Gomez; Writing – review & editing, Kourosh Kalantar-zadeh and Riccardo Frisenda..

# Supporting Information: Direct transformation of crystalline MoO₃ into few-layers MoS₂

**Felix Carrascoso *[1], Gabriel Sanchez-Santolino †,[1], Chun-wei Hsu [1,2], Norbert M. Nemes [3], Almudena Torres-Pardo [4], Patricia Gant [1], Federico J. Mompeán [1], Kourosh Kalantar-zadeh [5], José A. Alonso [1], Mar García-Hernández [1], Riccardo Frisenda [1], Andres Castellanos-Gomez *[1]**

1   Materials Science Factory. Instituto de Ciencia de Materiales de Madrid (ICMM-CSIC), Madrid, E-28049, Spain.
2   Kavli Institute of Nanoscience. Delft University of Technology. Delft 2600 GA, The Netherlands.
3   Departamento de Física de Materiales, Universidad Complutense de Madrid, E-28040 Madrid, Spain.
4   Departamento de Química Inorgánica, Facultad de Químicas, Universidad Complutense, 28040- Madrid, Spain.
5   School of Chemical Engineering, University of New South Wales, Kensington, NSW 2052, Australia.
*   Correspondence: felix.c@csic.es , andres.castellanos@csic.es

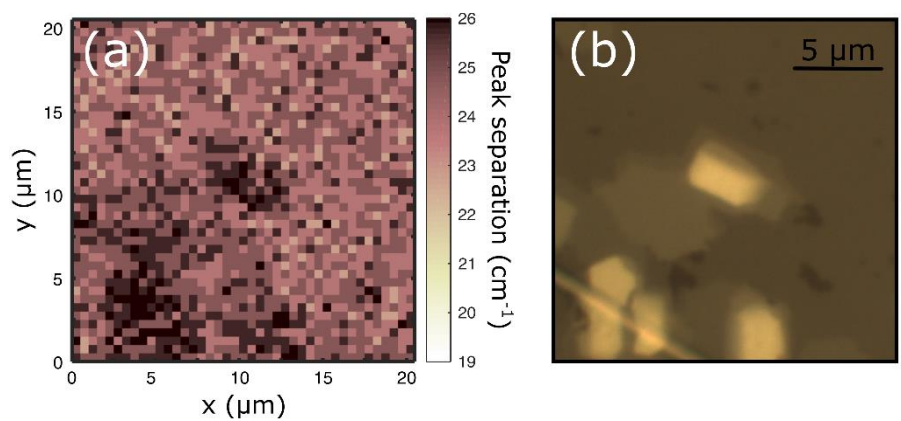

**Figure S1**: a) Raman map showing the difference between the $E^1_{2g}$ and $A_{1g}$ peaks. The map shows a large region of ~4 layers, according to the peak difference value, and a thicker region in the bottom left corner. b) Optical image of the same region studied in (a).

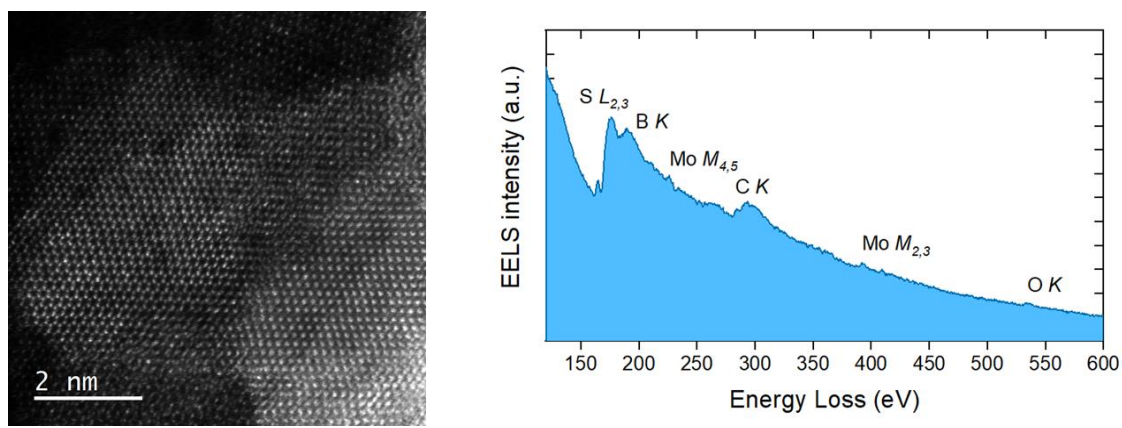

**Figure S2**: a) High magnification HAADF image of a MoS$_2$ thin film transferred over a holey Si$_3$N$_4$ membrane support. b) Electron energy loss spectra (EELS) acquired while scanning over the area in (a) for a total time of 20s.